\documentclass[twocolumn,showpacs,preprintnumbers,amsmath,amssymb]{revtex4}
\usepackage{graphicx}% Include figure files

\begin{document}

\preprint{APS/123-QED}
\title{Modulation of the nucleation rate pre-exponential in
a low-temperature Ising system}
\author{ Vitaly A. Shneidman }
\author{ Gelu M. Nita }

%uncomment 1 line below for revtex4 on aps
\affiliation{Department of Physics, New Jersey Institute of Technology, Newark, NJ 07102}
\date{\today}
\begin{abstract}
A metastable lattice gas with nearest-neighbor
interactions and  continuous-time dynamics  is studied using a 
generalized  Becker-D\"oring approach in the multidimensional space of cluster configurations. The
pre-exponential of
the
metastable state lifetime (inverse of  nucleation rate) is found to 
exhibit
distinct peaks at integer  values of the inverse supersaturation. Peaks are
unobservable (infinitely narrow) in the strict limit $T\rightarrow 0$,
but become detectable and eventually dominate at higher temperatures. 
\end{abstract}
\maketitle

In a general case, the nucleation-controlled lifetime of a  metastable
state can be written as
\begin{equation}
\tau= A\exp\left(W_\ast/T\right)
\label{volmer}
\end{equation}
 with
$W_\ast$ being the minimal work (free energy change) 
to form a critical nucleus
and the temperature $T$  measured in units of Boltzmann constant.
 The exponential term was
anticipated already in the earliest estimations \cite{class} of the
nucleation rate
$I\sim 1/\tau$, which further found an enormous amount of applications in systems
ranging from vapors
\cite{Abr74} to glass-forming \cite{Deb96} or  quantum
\cite{Leg84} liquids. The structure of the
pre-exponential, $A$, however, is not known in a general case. The difficulty
of its derivation, whether in an analytical, numerical or experimental study, 
stems from the dominant contribution of the exponential in eq.(\ref{volmer}),
with minor uncertainties in $W_\ast$ (say, due to inaccuracies in the measured 
interfacial tension) implying large, orders of magnitude, discrepancies in the values of $A$.

With this,  much attention is devoted to models
which can exhibit nucleation and which  are close to exact solvability.
Here one can  obtain an accurate expression for $W_\ast$, 
subsequently focusing on the pre-factor issue. 

One of the best known example is the two-dimensional nearest neighbor
Ising model, where 
metastability 
is achieved by orienting initially all spins one way (down)
 while non-zero magnetic 
 field $h$  prescribes an opposite (upward) orientation. Allowing spin
flips of
non-conserved 
 \cite{MetRosRos53Gla63} or conserved \cite{Kaw72} type adds the required
dynamics to the problem.
In a closely related lattice gas model the role of $h$ is played by 
supersaturation.
 
The pre-exponential in such systems attracts much attention both for the high- 
\cite{Fis67Lan71GunNikWal80GunMigSah83,StoBinSch72BinSta76,RikTomMiy94RicSidNov95,JCP99}
and low-temperature  \cite{Nov97,Nov02ParNov02,BovMan02} regions.
For $h\rightarrow 0$ the nucleus is macroscopic and its shape, as well as the value of $W_\ast$
can be obtained from the Wulff droplet construction 
\cite{RotWor81ZiaAvr82Zia86PRB01}.
Monte 
Carlo simulations are possible for $W_\ast\lesssim  10\;-\;15 \;T$ which, for small $h$, restricts such studies to the
aforementioned high-temperature region; transfer-matrix
approaches 
also are available here \cite{GunRikNov93GunRikNov94Rut01}. For larger
fields a 
straightforward Wulff construction may be inadequate, 
\cite{KotOli94}, but for $T\rightarrow 0$
 analytical treatment becomes possible due to  dominant contribution of low-energy
configurations. 
Technique of absorbing Markov chains  \cite{Nov97,Nov02ParNov02,BovMan02}
 also can be used for simulations in the low-temperature  region.

 Neves and Schonmann\cite{NevSch91} evaluated $W_\ast^0$, the
zero-temperature limit of $W_\ast$, 
obtaining the exponential part of the metastable lifetime. Their result is
insensitive to specifics of the dynamics. 
 Novotny \cite{Nov97} further showed that for  discrete-time
dynamics and a relatively large field,  the pre-exponential remains finite in the 
limit $T\rightarrow 0$, approaching a 
piece-wise constant function of $h$, pointing towards a discontinuity  at
an integer value of the inverse field.  
Similar features  will be observed at weaker fields as well
  \cite{BovMan02}. 
 Integer values of inverse field, however, were excluded from the
aforementioned rigorous mathematical treatments, leaving open questions
with regard to this
intriguing effect, especially in the physically more realistic case of
$T>0$.

The present Letter aims  to evaluate 
 the pre-exponential  at higher temperatures and  in a finite domain of
fields, 
spanning several integer values of $1/h$. This will clarify the nature of the
discontinuities and,  together with the available
 $W_\ast^0$, will
provide predictive expressions for $\tau(T,h)$ at $T>0$. We will show
that in contrast to intuitive expectation of
discontinuities spreading out in a standard, {\it tanh}-like fashion, 
they are replaced by sharp peaks which
persist, with  finite heights and  self-similar shapes, up to $T=0$.

A lattice gas model with continuous time
 dynamics  will be considered.  Specifically, the probability of creation of a particle on an 
empty site in an infinitesimal time interval $dt$ is taken as $\beta dt$, 
regardless of the surrounding; without restrictions, the time scale $\beta^{-1}$ can be taken as 1.
Alternatively, the 
annihilation probabilities are proportional to $dt\exp(-\Delta E/T)$, with 
$\Delta E$ being the energy change due to broken bonds and field
("supersaturation") which increases the energy by $2h$ when a particle is removed. This model, with various generalizations, 
is popular, e.g. in Monte Carlo simulations of the dynamic interface in 
crystallization problems - see, e.g. \cite{JacGilTem95JCG00} and
references therein.
It is expected that qualitatively the model also remains similar to the
discrete-time Glauber type  dynamics of
Refs.\cite{NevSch91,Nov97} 
(and the generalized Becker-D\"oring approach employed below bears certain
parallels with the technique of absorbing Markovs chains
\cite{Nov97,Nov02ParNov02}), 
although the dynamics-sensitive pre-exponentials 
will not be identical even at $T=0$.

 For a long time of the order of
$\tau$ the rare particles will form isolated clusters of various sizes and shapes (\emph{classes}), 
which will be distinguished by a running index, $i$. An empty site corresponds to $i=0$.
Cluster shapes will be considered identical (and thus belonging to the same 
class) if they can made such by rotation or reflection.
The key characteristics of each class are the numbers of particles, $s(i)$, the number of 
bonds $b(i)$ and the statistical weight $w_i\leq 8$. One can define the (quasi) equilibrium distribution
\begin{equation}
f_i^{eq}=w_i z^{2s(i)-b(i)}\delta^{s(i)}
\label{feq}
\end{equation}
with $
z=e^{-\varphi/T}$ and $\delta=z^{-2h}$
describing the temperature and field dependencies, respectively. $\varphi$ is the bond energy, subsequently taken as $1$
for simplicity of notations. In the $s, b$ space the function $ f_i^{eq}$ has a saddle point (for non-special fields - a single one \cite{BovMan02})
 and the 
corresponding value of $s$ determines the critical cluster number, $s_\ast$.
In a general case computer assistance is required in order to characterize all 
classes. Consistency of such predictions
 can be checked, e.g., against standard tables \cite{Dom60} for smaller $s$.

Once equilibrium properties are specified, one can introduce kinetic fluxes as a
multidimensional version of the classical approach \cite{class}, since 
in a low-temperature Ising system  growth or decay of a  cluster predominantly proceeds via random gain or loss of a single particle
\cite{MarMar84}.
If $\beta_{ik}dt$ is the probability 
to transform a cluster from class $i$ to class $k>i$ by adding a particle
[with $\beta{ik}=0$ if $s(k)\ne s(i)+1$], the corresponding flux  is given
by
\begin{equation}
I_{ik}=\beta_{ik}f_i^{eq}(v_i-v_k)\;,\;\;i<k
\label{Iik}
\end{equation}
with $v_i \equiv f_i/f_i^{eq}$ and $v_0 = 1$.
The Master Equation for the kinetic distributions $f_i$ takes the form
\begin{equation}
\frac{df_i}{dt}=\sum_{k=0}^{i-1}I_{ki}-\sum_{k=i+1}^{k_{\max}+1}I_{ik}
\label{dfdt}
\end{equation}
which automatically satisfies detailed balance.

For closing conditions,  absorbing states are placed
at all classes $k$ with $s(k)= s_{\max} +1$. 
Equivalently, all those absorbing states can be
combined in a single absorbing class $k_{\max}+1$.

Due to an exponentially long lifetime, one can neglect 
the depletion of 
empty sites (for which, otherwise, an integral conservation law \cite{Pen97} should be 
employed instead of  $v_0\equiv 1$).
 With this, eqs. $(\ref{Iik})$, $(\ref{dfdt})$ can be solved in the
steady-state 
approximation; transient effects \cite{gelu_rem} also can be neglected
here. 

Introducing $b_{ik}=\beta_{ik}f_i^{eq}+\beta_{ki}f_k^{eq}$ ($0\le i,k\le
k_{\max}+1$) and 
\begin{equation}
M_{ik}=b_{ik}
-\delta_{ik}\sum_{l=0}^{k_{max}+1}b_{il}\;,\;\;1\le i,k\le k_{\max}
\label{mik}
\end{equation}
% ($1\le i,k\le k_{\max}$),
 one can show that the steady-state distributions $(v_1,\;v_2,\;\ldots)$
correspond to the first column of the matrix $-\hat M^{-1}$ (since only 
 class $1$, with single-particle clusters, is connected to empty
sites).  The total
 flux $I$
coincides with $I_{01}$ where branching of paths does
not yet occur. This gives
 \begin{equation}
I=(\hat M^{-1})_{11}+1
\label{Imhat}
\end{equation}

For a single nucleation path, which leads to a tri-diagonal structure of
the matrix $\hat M$, one recovers the classical result by Farkas
\cite{class}
$I^{-1}=b_{01}^{-1}+b_{12}^{-1}+\ldots$.
Otherwise, the actual evaluation of $I$ via eq. (\ref{Imhat})  is limited
by 
one's ability to obtain all classes and transition rates $\beta_{ik}$ for a sufficiently large
$s_{\max}$, and the ability to inverse analytically a large matrix $\hat{M}$.
At present, we were able to proceed up to $s_{\max}=9$ (1818 classes
representing a total of
13702 cluster configurations) which allows us to consider fields $h>1/6$ 
 with the critical number $s_\ast \leq 7$. 
A full exact  expression for $\tau$  can be surveyed by a human
eye only for more modest values of $s_{max}$, which implies a relatively small critical cluster 
(larger fields). 
For example, for $s_{\max}= 4$  kinetics is determined by 9 distinct classes with a total of 28 shapes 
(see, e.g. Fig. $2$ in Ref.\cite{JCP99}).
Transition rates are easy to obtain (say, there are two ways a 3-particle "minus" shaped cluster can turn
into 4-particle "T" shaped one,  four ways it can turn into an "L" shaped one, etc.).
The result which follows from eq.(\ref{Imhat}), is expressed as a rational function of $z$ and $\delta$
\begin{equation}
\tau_{4}={P(\delta,z)}/{Q(\delta,z)}
\label{4spin}
\end{equation}
with the subscript indicating the value of $s_{\max}$, and  polynomials $P$ and $Q$ given by
\begin{eqnarray}
\label{4full}
P(\delta,z)&=& 384 + 210000 \delta^9 z^9 + 16 \delta (48 + 185 z) + \\
\nonumber
      && 4 \delta^2 z (1576 + 2655 z) +8 \delta^3 z^2 (3081 + 3230 z) +\\
\nonumber
      && 2500 \delta^8 z^7 (21 + 250 z + 36 z^2)+ \\
\nonumber
      && 250 \delta^7 z^6 (695 + 2650 z + 1036 z^2) +\\
\nonumber
      && \delta^4 z^3 (65740 + 57797 z +15360 z^2) +\\
\nonumber
      && 5 \delta^5 z^4 (28574 + 28155 z + 19680 z^2) + \\
\nonumber
      && 5 \delta^6 z^5 (43375 + 70546z + 50000 z^2)\\
\nonumber
Q(\delta,z)&=& 8 \delta^4 z^4 (384 + 80 (24 + 85 \delta) z +\\
\nonumber
      && 250 \delta^2 (25 + 64 \delta) z^3 +125 \delta^3 (259 + 625 \delta) z^4 + \\
\nonumber
      && 20 \delta (615 + 1753 \delta) z^2 + 3750 \delta^4 (3 + 7 \delta) z^5)
\end{eqnarray}  

Eq.(\ref{4spin}) is expected to be accurate in strong fields, $h\gtrsim 1/2$, with rather relaxed restrictions on temperature
since all cluster configurations at $s\le 4$ are taken into account
(although, for higher $T$ eventual destruction of the steady-state   due to neglected  cluster interactions should be kept in 
mind \cite{PRB99}). More consistently, this result should be treated asymptotically for $z\rightarrow 0$ and $\delta\rightarrow\infty$ with certain combinations of powers of $z$ and $\delta$ remaining finite, depending
on the interval of field.
 
\begin{figure}
\scalebox{0.8}{
        \rotatebox{-0}{
\includegraphics{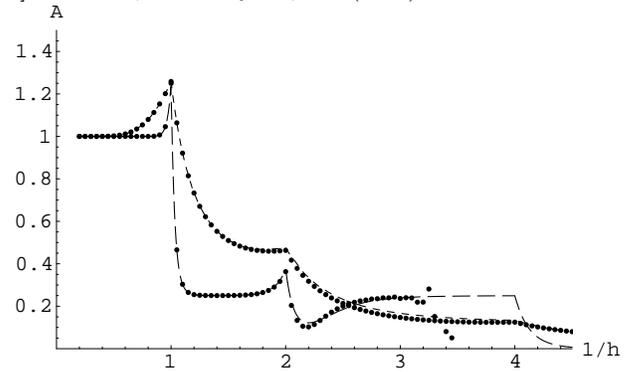}
}}
\caption{
The pre-exponential of eq. (\ref{volmer}).
Lines -  the 4-particle approximation, eq.(\ref{4spin}), at
a higher temperature, $z=10^{-1}$ (short dashed) and at a low temperature,
$z=10^{-7}$ (long dashed). Points - numerical results for $s_{max}=9$ and $r=2$.
}
\label{fig1}

\end{figure}

In order to isolate the pre-exponential,  eq.(\ref{4spin}) should be multiplied by 
$\exp(-W_\ast/T)=z^{W_\ast}$. In principle, an "observable"
is $\tau$ itself, rather than $A$ or $W_\ast$ taken separately. 
To avoid ambiguity, the value of $W_\ast$ will be associated
with its zero-temperature limit, $W_\ast^0$ \cite{NevSch91}, with all temperature-dependent corrections being 
in the pre-exponential; for $h> 1$ the barrier will be taken as zero.  The
function $W_\ast^0(h)$ has a 
piece-wise linear structure, and $\exp(-W_\ast/T)$
is reduced to a product of integer powers of $z$ and $\delta$: $1$ for $h\ge 1$, $z^2\delta$ for $1/2\le h<1$,
$z^4\delta^3$ for $1/4 \le h< 1/2$, etc. The resulting $A(h)$ is shown by a dashed line in Fig. \ref{fig1}   
  where numerical results,
given as filled circles, were obtained for a much larger $s_{\max}$ and
can be treated as "exact" in the present context.
The case $T=0$ would correspond to a piece-wise constant structure of $A$,
similar to Ref. \cite{Nov97} but with different constants and
an additional "excluded singularity" at $h=1/2$: $A=1$ for $h>1$, $A=1/4$
for $1>h>1/4$ ($h\ne 1/2$), $A=1/16$ for $1/4>h>1/6$, etc. (These
numbers can be deduced from the lowest energy path -see below-,
serving as a checkpoint for more more elaborate expressions). This
limit, however, becomes
apparent only at a very low temperature, $z= 10^{-7}$.

For a larger cut-off, simplifications of analytics can be achieved due to the dominant contribution of low-energy configurations.
Among all classes $k$ of clusters with the same  $s(k)$, one can select only
those which have a sufficiently large number of bonds: $b_s(k)\geq b_{max,s(k)}-r$,
where $b_{max,s}$ is the number of bonds in the most compact cluster for a given $s$. 
An integer  parameter $r$ indicates how close a cluster should be to the most compact configuration
in order to be included in the kinetics. For sufficiently
large $r$  ($r=4$ for $s_{max}=9$) all configurations are recovered.
Alternatively, $r=0$ corresponds to the lowest energy path description, which is the closest to the kinetic part of the
conventional one-dimensional random walk  approach to nucleation \cite{class}, although with microscopic rather than phenomenological coefficients.
In addition, branching of paths is added starting from $s=7$.
Already in the $r=0$ approximation  peaks at integer $1/2h$ will appear in the pre-exponential, although
 one needs to include $r\ge 1$ for  correct evaluation of their heights.

For the case $r =1$ and $z\ll 1$ the pre-exponential $A(h)$ can be described 
analytically  in  restricted domains of fields, 
the most interesting being those near the peaks (general expressions for $A(h)$ are also 
 available, but are useless due to their size).

Introducing a {\it finite} combination
\begin{equation}
\label{y_z}
y=\delta z^{1/n}
\end{equation}
with $n=1,2,\ldots$ determining a 
corresponding peak, one can perform analytical 
expansions of $1/I$ in fractional powers of $z$.
Symbolic computations with \emph{Mathematica} were used here. 

\begin{figure}
\scalebox{0.8}{
        \rotatebox{-0}{
\includegraphics{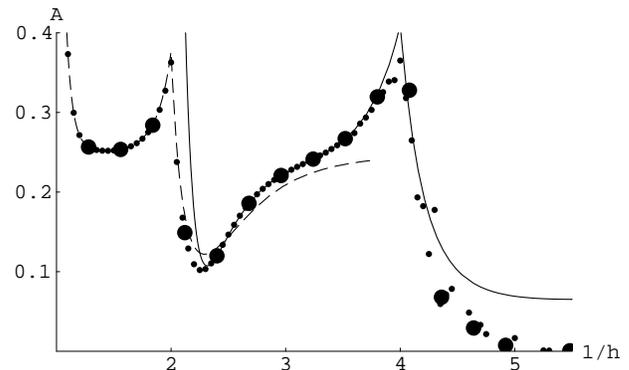}
}}
       \caption{
"Fine structure" of the two peaks near $h=1/2$ and $h=1/4$ for $z=10^{-5}$. 
Solid line- the 9-particle approximation for $r=1$, eq.(\ref{tau9}). 
Dashed line- the full 4-particle approximation, eq.(\ref{4spin}).
Symbols - numerical data for $s_{max}=9$ and $r=2$ (small circles) 
and $r=4$ (large circles), respectively. 
}
\label{fig2}
\end{figure}

For  $n=2$  one obtains
\begin{equation}
\label{tau9}
\tau_{9}=\frac{1}{z^{5/2}}\frac{T_1(y)}{8y^7 T(y)} -
\frac{1}{z^2}\frac{T_2(y)}{16y^8 T^2(y)} +
 \frac{1}{z^{3/2}}\frac{T_3(y)}{672y^9 T^3(y)}+\ldots
\end{equation}
with the coefficients in this $s_{\max}=9$ approximation given by 
\begin{eqnarray}
T(y)&=& 8 + 63y^2\\
\nonumber
T_1(y)&=& 4 + 29y^2 + 79y^4 + 126y^6\\
\nonumber
T_2(y)&=&-208 - 1432y^2 + 3461y^4 + 49855y^6 +\\
\nonumber
      &&  89649y^8 + 87318y^{10}\\
\nonumber
T_3(y)&=& 89152 - 297840y^2 - 13174644y^4 - 62801445y^6 +\\
\nonumber
      && 146767614y^8 +1284356493y^{10} + \\
\nonumber
      && 957680010y^{12} + 556604622y^{14}
\label{T9}
\end{eqnarray}                             
The approximation works accurately in the vicinity of $h=1/4$, describing
the rather complex near- and  off-peak behavior - see Fig. \ref{fig2}. 
The coefficient of $z^{-5/2}$ in eq.(\ref{tau9}) multiplied, respectively 
by $y^3$ at $h>1/4$ or by $y^7$ at $h<1/4$,
determines the scaling structure of the peak in the limit $T\rightarrow 0$, if the difference $h-1/4$ scales together with temperature.
The structure of the neighboring peak at $h=1/2$ ($n=1$) follows the
4-particle approximation, eq.(\ref{4spin}), with
$\delta=y/z$ and $z\rightarrow 0$.

An important question is the sensitivity of the results to variations in $r$
and $s_{max}$. A smaller $r=0$ 
can reproduce the coefficient of $z^{-5/2}$ in eq.(\ref{tau9}) in the
limits $y\rightarrow\infty$ or $y\rightarrow 0$
(i.e. on both sides of the peak for $T\rightarrow 0$), but will not give the 
proper peak height for $y=1$, or  scaling for finite $y$, or the correct
higher-order terms.
 Cases with larger
$r\ge 2$ presently  could be studied only numerically  and are shown by symbols in Fig. \ref{fig2}.
At small $h$ scatter appears in data, indicating the limits of numerical accuracy for very small $z$.
There is no detectable difference with the analytical approximations in the 
regions of their validity for fields up to $h\gtrsim 1/4$. 
An similar expansion in $z$ for
 $r=1$ and $s_{max}=8$ also was performed,
leading to a rather different structure of the $y$-dependent
polynomials. The
 first coefficients in the $z$-expansion 
are nevertheless numerically close
for stronger fields $ h\ge 1/4$ in the vicinity of the peak.
So are the 
heights of the peak    
 given, respectively by
$0.4167-2.92z^{1/2}+\ldots$
and $0.4190-2.83z^{1/2}+\ldots$ in the 8-and 9-particle approximations.
On the other hand, unlike the 9-particle case,  $s_{\max}=8$ 
 does not yield a proper $T\rightarrow 0$ limit for weaker fields $h<1/4$
since
the boundary here is too close to the critical size $s_\ast=7$.

In summary, for a moderate field (supersaturation) the metastable 
state lifetime of a supersaturated lattice gas has been evaluated for
$T\ll T_c$. 
The main result is the pre-exponential which, for the first time,
was evaluated analytically beyond the zero-temperature limit, and which
exhibits distinct peaks as a function of field.
One can anticipate that similar peaks (which appear due to competition
of several "critical sizes") also will be observed in systems other than 
 nearest-neighbor Ising
models with non-conserved dynamics, whenever the nucleation barrier has a
well-defined zero-temperature limit and the critical nucleus
contains a reasonably small number of particles. 

\acknowledgments 
Authors are grateful to Mark Novotny for useful correspondence and 
comments on the manuscript.

%uncomment below for preprint version
\newpage

%\bibliography{vit,my,kram,rip,string,remarks,nest1,ising}

%uncomment below for preprint version
%\newpage

\end{document}